# A Study on Delay Assessment for Heterogenous Traffic in VANET

Shama Siddiqui* [1,2], Indrakshi Dey[2]

[1] *DHA Suffa University,* Karachi, Pakistan
email: shamasid@hotmail.com
0000-0002-3547-0307

[2] *Walton Institute for Information and Communication Systems Science*
South East Technological University, Waterford, Ireland
email: indrakshi.dey@waltoninstitute.ie
0000-0001-9669-6417

*Abstract* —Vehicular Ad hoc Networks (VANETs) comprise of multi-priority heterogenous nodes, both stationary and/or mobile. The data generated by these nodes may include messages relating to information, safety, entertainment, traffic management and emergency alerts. The data in the network needs differentiated service based on the priority/urgency. Media Access Control (MAC) protocols hold a significant value for managing the data priority. This paper studies a comparison of 802.11p which is a standard PHY and MAC protocol for VANET with a fragmentation-based protocol, FROG-MAC. The major design principle of 802.11-p is to allow direct Vehicle-to-Vehicle (V2V) and Vehicle-to-Infrastructure (V2I) communication without association, using Enhanced Distributed Channel Access (EDCA) to prioritize safety-critical messages. However, if non-critical messages already start to transmit, the nodes with critical data have to wait. FROG-MAC reduces this delay by transmitting normal packets in fragments with short pauses between them, allowing urgent packets to access the channel during these intervals. Simulations have been performed to assess the delay and throughput for high and low priority data. We report that FROG-MAC improves both the performance parameters due to offering an early channel access to the emergency traffic.

## Introduction

Lately, the protocols focusing on providing efficient, reliable and secure wireless communication in the vehicular environment have been significantly advanced. Many types of data dealing with information, safety, entertainment, emergency alerts, and other messages are transmitted for Vehicle-to-Vehicle or Vehicle-to-Everything (V2X) links [1]. It is crucial to offer the required level of service in terms of priority management, delay, and packet loss for each node to ensure smart urban traffic management [2]. Additionally, these protocols must address the challenges of high mobility, dynamic network topology, hardware limitations, channel congestion and variable traffic density in vehicular environments [3]. Developing solutions that optimize channel utilization and prioritize critical data transmission is vital for enhancing the performance of VANET.

The United States Federal Communications Commission (FCC) has designated the 5.850–5.925 GHz frequency band to facilitate VANET communications [4]. This 5.9 GHz band is split into seven non-overlapping channels, each with a bandwidth of 10 MHz. Communication within VANETs is standardized through the use of Dedicated Short-Range Communication (DSRC) based on IEEE 802.11-p and Wireless Access in Vehicular Environments (WAVE) [5]; these technologies ensure high data transfer rates, reliable connectivity, and low latency. Moreover, CSMA/CA protocols are employed at the MAC layer, within the WAVE standard to handle the challenges posed by high mobility and dynamic network topologies [6].

IEEE 802.11-p has been designed specifically for VANET, by modifying the Basic Service Set (BSS) mechanism of 802.11 to operate without requiring association or authentication between nodes, allowing faster and more flexible communication in high-mobility environments [7]. 802.11-p uses Outside the Context of a BSS (OCB) mode, enabling direct communication between vehicles or between vehicles and roadside units without the need for a central access point [8]. Prioritized access in 802.11-p is achieved through the Enhanced Distributed Channel Access (EDCA) mechanism [9], further discussed in section II. Despite the use of EDCA, 802.11-p can cause a higher delay for urgent traffic, since it does not allow any chance for this traffic to transmit if the traffic of lower priority has already started to transmit.

In this work, we study the performance comparison of 802.11-p with FROG-MAC, which is designed for prioritized heterogenous traffic [10]. FROG-MAC offers a pseudo-preemptive priority to the nodes with urgent traffic by fragmenting the lower priority packets and introducing pauses between two consecutive fragments. This approach ensures that if an urgent packet is generated while the lower priority data is already being transmitted, the node will get a chance to transmit a Request to Send (RTS) frame during the next pause [11]; conventionally, the nodes can only begin their urgent transmission process once the lower priority packets complete transmission, which increases the delay and hence, limits the performance of protocols for smart traffic environments. Therefore, FROG-MAC offers an opportunity of early transmission, promising better performance in terms of delay and reliability for future networks. Further details of FROG-MAC illustrating the use case of heterogenous traffic will follow in section II. To the best of our knowledge, the comparison of IEEE 802.11-p has not been conducted with a fragmentation-based approach in the past. The major contribution of this work are as follows:

- To evaluate via simulation IEEE 802.11p versus FROG-MAC under mixed high-and low-priority VANET traffic, showing that FROG-MAC achieves lower delay and higher throughput.

- To demonstrate that FROG-MAC's fragmentation mechanism provides pseudo-preemptive access for urgent traffic.

- To provide the first known comparison of IEEE 802.11p with a fragmentation-based MAC protocol in a VANET context.

Rest of this paper has been organized as follows: section II briefly summarizes the relevant literature; section III describes the experimental settings; section IV details the results and finally, section V concludes the work and suggest future directions.

## Relevant Work

With the advent of smart city concepts, the vehicular networks and associated protocols have gained wide popularity lately; this is because efficiently designed MAC protocols for VANET will lead to better road safety and overall traffic management. In this context, researchers have proposed various schemes for achieving the objective of optimizing network performance; for example, the use of adaptive contention window [12], use of multi-channels [13], prioritization of critical data [14], and clustering techniques [15]. Additionally, mechanisms such as predictive modeling for traffic patterns [16], and integration of cross-layer designs [17] have been explored to enhance reliability and minimize latency. We briefly discuss some of them in this section:

A number of multi-channel MAC protocols have been proposed for VANET, where commonly, one channel is dedicated for transmitting the control information and others are allocated for sending the messages of different priorities each. Many variants of this approach are found in literature; for example, in [18], a synchronized protocol has been proposed which does not employ a dedicated control channel. Instead, for each node, a channel table is established using a straightforward handshaking mechanism, which records the status of channels and neighboring nodes; this enables each node to determine both the priority of its own data and the priorities of its neighbors. The sender nodes initiate synchronization by assigning priorities, allowing the method to identify a shared channel directly or indirectly if both transmitter and receiver nodes have one in common. Furthermore, the scheme incorporates preemptive data transmission to minimize the frequency of channel switch-overs during data transfer. Although multi-channel protocols make it easier to deal with heterogenous data of multiple priorities, there are hardware limitations as well as challenges related to channel coordination and synchronization; these constraints can lead to increased complexity in managing multiple channels [19].

The standard protocol IEEE 802.11-p offers significant advantage to VANET in comparison to the conventional 802.11, including support for high-speed mobility and direct communication through its OCB mode [9]. It uses 7 channels, each with a bandwidth of 10 MHz, allocated within the 5.9 GHz frequency band. These channels include one control channel (CCH) and six service channels (SCHs): CCH is used for exchanging safety-critical information, and SCHs are used for transmitting data generated by non-safety applications, such as infotainment and traffic management; this channel structure allows 802.11p to efficiently prioritize urgent safety communications while supporting other vehicular applications.

Additionally, 802.11-p reduces communication delays by eliminating the need for association and authentication processes, which makes it more fit for real-time vehicular applications [20]. Moreover, the protocol also offers prioritized access through EDCA by categorizing data into four access categories (ACs): voice, video, best effort, and background. Each AC is assigned specific contention window (CW) sizes and arbitration interframe spaces (AIFS), ensuring that high-priority traffic, such as safety-critical messages, experiences shorter waiting times and faster access to the channel. The use of AIFS for different traffic priorities has been illustrated in figure 1 [21]. By reducing backoff times and prioritizing critical transmissions, DSRC effectively manages channel access in less congested environments, making it suitable for real-time vehicular safety applications. Despite several unique features, 802.11-p does not offer a chance of transmission to higher priority traffic if the lower priority has already started.

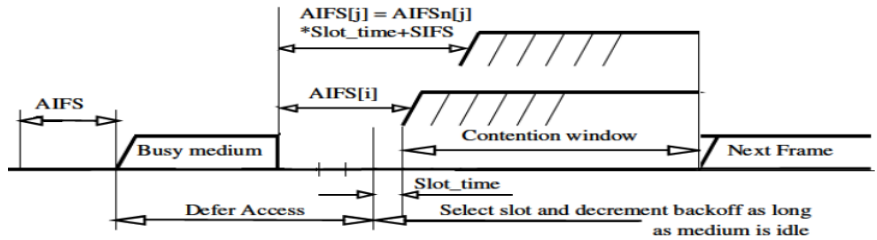

*Figure 1: EDCA Mechanism for Priority Assignment in 802.11-p.*

DSRC (developed on the principles of IEEE 802.11-p) and LTE-V2X have been compared in [5]; both protocols differ significantly in their vehicular communication mechanisms: While DSRC relies on a distributed contention-based mechanism using the EDCA, LTE-V2X employs a sensing-based semi-persistent scheduling (SPS) approach. Although 802.11-p appears a good choice for lesser congested environments as discussed above, its reliance on fixed backoff parameters and decentralized control can lead to scalability issues in highly dynamic or dense networks. On the other hand, LTE-V2X is better suited for handling priority traffic, especially in scenarios with high mobility and larger communication ranges; its dynamic scheduling and resource allocation provide more flexibility and scalability compared to DSRC's fixed contention-based approach.

In comparison with 802.11-p, we study FROG-MAC in this work [11]. FROG-MAC has been designed with the goal of reducing delay of heterogenous traffic, by introducing the novel feature of channel interruption. Figure 2 shows the basic operation of FROG-MAC, where it can be seen that the lower priority/normal traffic is sent in fragments, which allow the higher priority nodes to send a RTS frame between two fragments of lower priority. In the absence of this interruption mechanism, urgent data could only be transmitted after the normal data had been fully transmitted. Therefore, FROG-MAC attempts to improve the delay for critical traffic, such as safety messages or emergency alerts in the VANET. The detailed flow operation of FROG-MAC's sender node has also been shown in Figure 3.

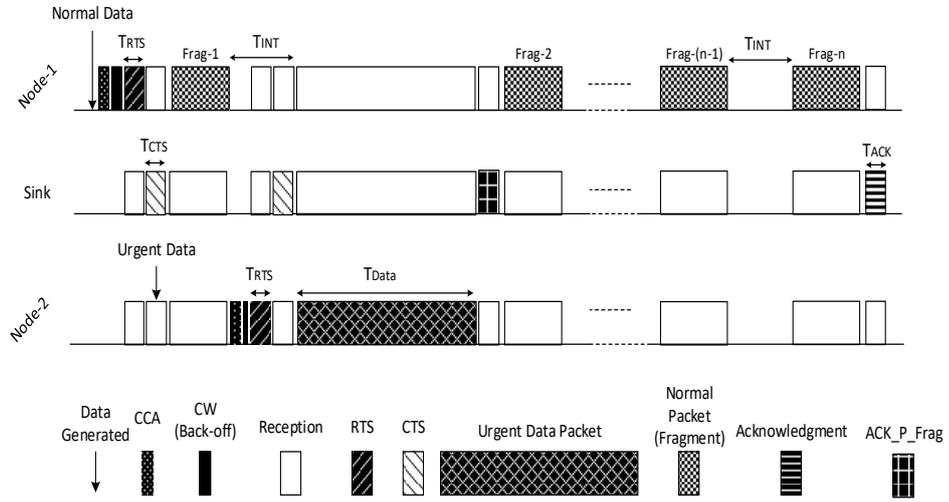

Figure 2: Basic Operation of FROG-MAC

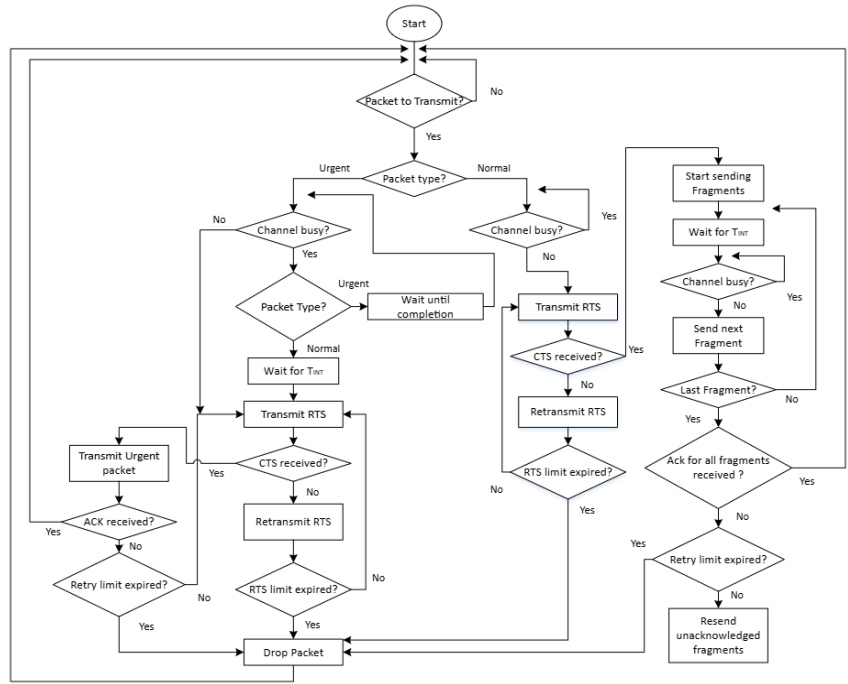

Figure 3: Flow Diagram for Sender's Operation in FROG-MAC

## Experimental Settings

We assumed 11 nodes arranged in star topology for comparing the performance of 802.11-p and FROG-MAC; this setting illustrates the vehicular nodes (source) sending messages to RSU (sink). We considered traffic of two priorities, where AC0, AC1 and AC2 as defined by 802.11-p have been considered normal, and AC3, as urgent. Here, the normal messages may represent routine updates, such as vehicle speed, location, or environmental data like temperature and air quality, which are used for traffic management or monitoring. In contrast, the urgent packets could signify situations such as possible collisions, emergency brakes, or notifications about fast approaching obstacles [22]. Clearly, AC3 should be prioritized in order to ensure that the overall purpose of VANET is served and safety could be guaranteed.

*Table 1: Simulation Settings*

| Simulation Parameter | Value |
| --- | --- |
| Simulation Time | 1000 Sec |
| Coverage Area | 1000 X 1000 m |
| Transmission Power | 23 dBm |
| Contention Window Size | Non-overlapping 3 to 15 for urgent traffic 16 to 1023 for normal traffic |
| Network Size | 2 to 11 |
| AIFS for Urgent Traffic | 18 μsec |
| AIFS for Normal Traffic | 27 μsec |
| Urgent Traffic Generation Interval | 2 sec (Poisson) |
| Normal traffic Generation Interval | 200 msec (CBR) |
| Size of Data Packet | 127 Bytes |
| Length of Total Payload | 121 Bytes |
| Fragment size (payload length in each fragment) | Varying (2 to 121Bytes) |
| No. of Fragments | Varying (1 to 61) |
| Size of RTS/CTS Packets | 5 Bytes |
| Size of ACK Packet | 5 Bytes |
| Size of SACK/NACK Packets | 6 Bytes |
| Size of ACK_P_Frag (A packet that signifies pending fragments) | 8 Bytes |
| Delay for Transmitting each byte | 32 μsec |
| Interruptible Period $T_{INT}$ | 0.6 msec |

## Results and Discussion

Contiki was used to simulate the Delay and throughput performance for 802.11-p and FROG-MAC for the two types of traffic studied: normal and urgent. Since the core contribution of FROG-MAC is fragmentation, the simulations were run for fragment sizes 2 and 16, to also visualize the impact of fragmentation on each type of traffic. On the other hand, there is no fragmentation in 802.11-p, so it does not have any fragmentation-related parameter to configure; instead we have considered the protocol as an example of non-fragmented scheme. Each experiment was performed 5 times, and the results reported in this section have been presented with 95% confidence.

### *Average Delay*

Average Delay for 802.11-p and FROG-MAC has been compared in figures 4(a) and (b). In fig. 4(a), the fragment size was kept as 2 for the FROG-MAC, whereas it was varied to 16 in fig. 4(b). As mentioned before, since 802.11-p does not apply fragmentation, there is no impact on its results.

As seen in figure 4, average delay increases with the increasing number of nodes; this is because the number of collisions and retransmissions increases with the higher number of contending nodes on the network. In fig. 4(a), the delay for normal traffic is the highest for FROG-MAC, whereas delay of urgent traffic is the lowest for the same protocol. On the other hand, the delay of normal traffic has been found to be better for 802.11-p due to its design, however, the performance of FROG-MAC still remains superior for urgent traffic. This observation has been recorded because for the lower priority traffic, FROG-MAC intentionally fragments the packets and includes pauses, whereas no such scheme is implemented in 802.11-p. On the other hand, despite 802.11-p attempts at facilitating the higher priority traffic via adjusting CW and AIFS accordingly, it is clear that once the normal traffic begins transmission, it cannot be interrupted. Hence, FROG-MAC which allows interruption of active transmission of lower priority data, appears to be the better choice for urgent traffic.

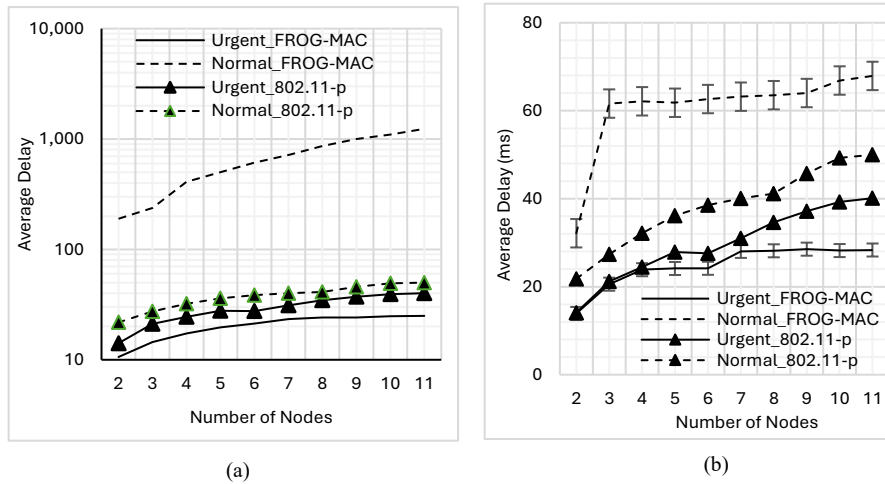

(a)                                        (b)

*Figure 4: Average Delay: (a) Fragment size=2, (b) Fragment size=16*

Similar concept is reported in fig. 4(b), where FROG-MAC again appears the worst for normal and best for urgent; the trends show difference in values as compared to fig. 4(a) because as the fragment size has been increased to 16 now, the urgent packets will need to wait a little higher before they can interrupt the lower priority fragments.

### *Throughput*

Figure 5 compares the throughput performance of the two protocols 802.11-p and FROG-MAC. In fig. 5(a), the fragment size is 16, whereas it is 2 in fig. 5(b). The trends show that, initially, both protocols achieve maximum throughput for both normal and urgent traffic. However, as traffic load increases, the throughput for normal traffic in FROG-MAC decreases to a minimum, while the throughput for urgent traffic reaches its maximum. Another interesting trend observed in fig. 5(a) and (b) is that for the fragment size 16, the maximum throughput for urgent traffic is almost same for the FROG-MAC and 802.11-p at the higher traffic load. On the other hand, for fragment size 2, the throughput of FROG-MAC is higher for urgent traffic,  as it gets more opportunity of quicker transmission.

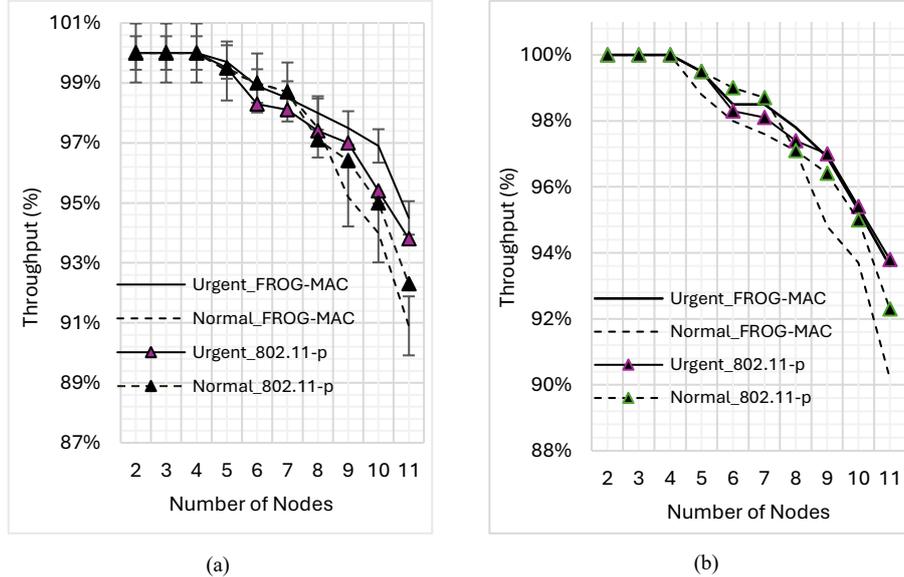

(a)                              (b)

*Figure 5: Throughput Evaluation: (a) Fragment size=2, (b) Fragment size=16*

## Conclusion and Future Work

VANET requires efficient MAC protocol design to guarantee the safety and urban traffic management. This paper evaluated 802.11p versus FROG-MAC for average delay and throughput. 802.11p improves the conventional MAC by eliminating BSS requirements and prioritizing emergency messages on the CCH. FROG-MAC, by fragmenting lower priority packets, gives urgent traffic earlier transmission opportunities and, as shown in our simulations, yields lower delays under moderate node densities.

The evaluation was conducted in a simulated environment with simplified mobility patterns and only measured average delay and throughput, so real-world RF effects, extreme congestion, packet delivery ratios, and fragmentation, reassembly overhead were not captured. Future work should include testing FROG-MAC under high density vehicular scenarios to assess scalability and reliability, and evaluating it alongside emerging standards, such as 5G NR V2X or 6G, for ultra reliable, low latency communication.

## Acknowledgment

This contribution is supported by HORIZON-MSCA-2022-SE-01-01 project COALESCE under Grant Number 10113073.